%% file: preprint.tex
\documentclass[journal, onecolumn, draftclsnofoot]{IEEEtran}
\usepackage{cite}
\usepackage{amsmath,amssymb,amsfonts}
\usepackage{algorithmic}
\usepackage{graphicx}
\usepackage{textcomp}

\include{macros}

\def\BibTeX{{\rm B\kern-.05em{\sc i\kern-.025em b}\kern-.08em
    T\kern-.1667em\lower.7ex\hbox{E}\kern-.125emX}}
    
\begin{document}
\title{Charge--Trapping--Induced Compensation of the Ferroelectric Polarization in FTJs: optimal conditions for a synaptic device operation}
\author{R. Fontanini, M. Segatto, K. S. Nair, M. Holzer, 
F. Driussi, \IEEEmembership{Member, IEEE}, I. H\"ausler, C. T. Koch, C. Dubourdieu, \IEEEmembership{Senior Member, IEEE}, V. Deshpande and D. Esseni, \IEEEmembership{Fellow, IEEE}
\thanks{This work was supported by the European Union through the BeFerroSynaptic project, Grant 871737.}
\thanks{R. Fontanini (fontanini.riccardo@spes.uniud.it), M. Segatto, F. Driussi and D. Esseni are with the Polytechnic Department of Engineering and Architecture (DPIA), University of Udine, via delle Scienze 206, Udine, Italy. }
\thanks{K. S. Nair, M. Holzer and C. Dubourdieu are with Helmholtz--Zentrum Berlin, Hahn--Meitner--Platz 1, Berlin, Germany and with Freie Universit\"at Berlin, Physical Chemistry, Arnimallee 22, 14195, Berlin, Germany.}
\thanks{V. Deshpande is with Helmholtz--Zentrum Berlin, Hahn--Meitner--Platz 1, Berlin, Germany.}
\thanks{I. H\"ausler and C. T. Koch are with 
Institut f\"ur Physik, Humboldt-Universit\"at zu Berlin, Berlin, Germany.}
\thanks{R. Fontanini, M. Segatto, K. S. Nair and M. Holzer contributed equally to this work.}}

\maketitle

\begin{abstract}
In this work, we present a clear evidence, based on numerical simulations and experiments, that the polarization compensation due to trapped charge strongly influences the ON/OFF ratio in Hf$_{0.5}$Zr$_{0.5}$O$_{2}$ (HZO) based Ferroelectric Tunnel Junctions (FTJs). Furthermore, we identify and explain compensation conditions that enable an optimal operation of FTJs. Our results provide both key physical insight and design guidelines for the operation of FTJs as multi--level synaptic devices.
\end{abstract}


\textbf{© 2023 IEEE.  Personal use of this material is permitted.  Permission from IEEE must be obtained for all other uses, in any current or future media, including reprinting/republishing this material for advertising or promotional purposes, creating new collective works, for resale or redistribution to servers or lists, or reuse of any copyrighted component of this work in other works.}

\section{Introduction}
\label{sec:introduction}
\IEEEPARstart{T}{he} demand for energy efficient hardware platforms for artificial intelligence has raised a growing interest for memristive devices capable of implementing non volatile, multi--level conductance \cite{Yu_IEEE_Proc2018}.
Prominent applications of such synaptic devices are crossbar arrays for artificial deep neural networks \cite{Ambrogio_Nature2019}, and hybrid memristive--CMOS circuits for spike--based neuromorphic computing \cite{Chicca_APL2020}.

\begin{figure}[!t]
\centerline{\includegraphics[width=\columnwidth]{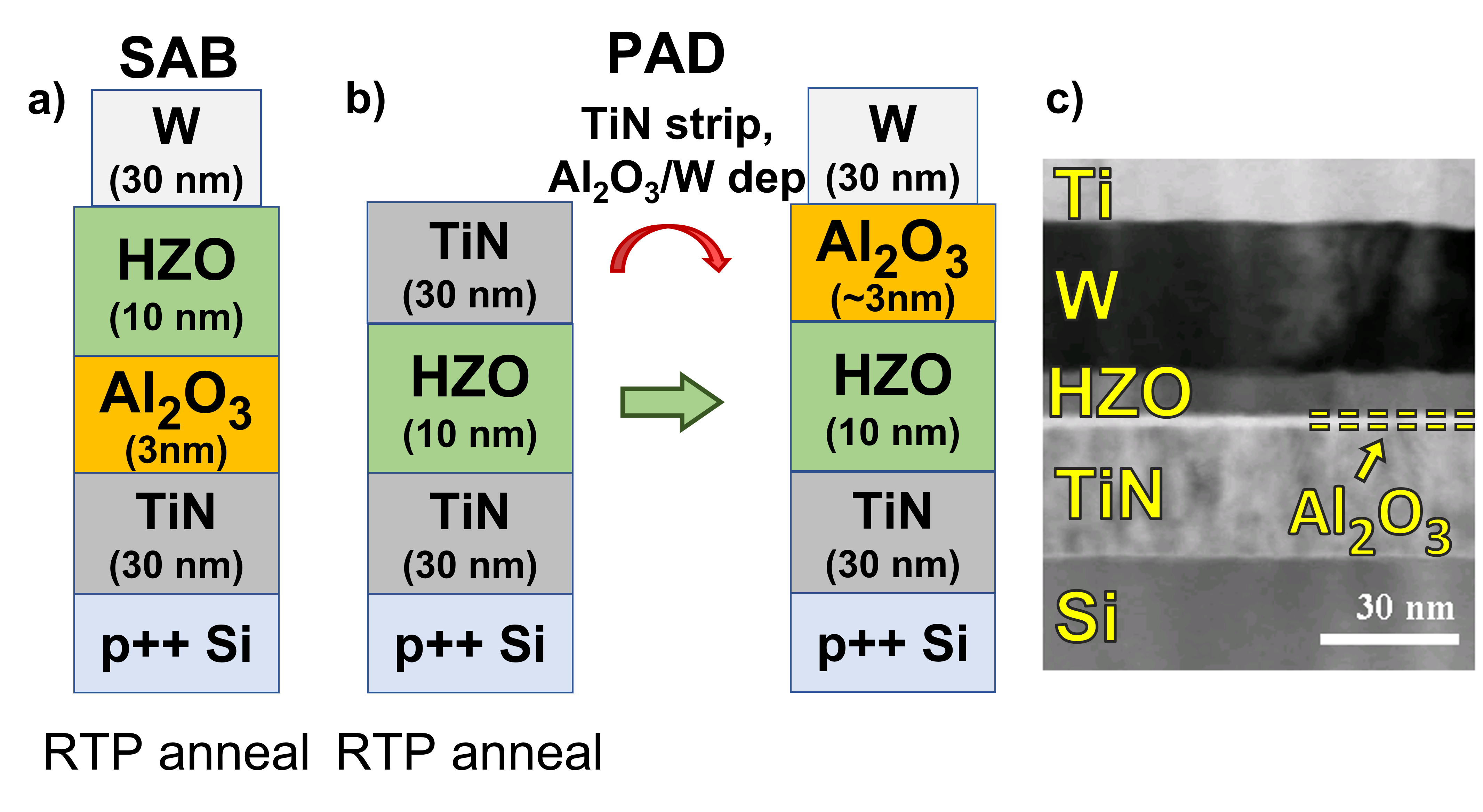}}
\caption{a), b) Sketches of the Stack Anneal--Bottom (SAB) FTJ and of the Post Anneal--Deposition (PAD) device, respectively. The main difference between SAB and PAD is the annealing process undergone by the Al$_2$O$_3$. Devices have a top electrode area of 95$\times$95~$\mu$m$^2$. c) Cross section TEM image of the SAB FTJ stack 
in this work.}
\label{Fig:FTJ_sketch}
\end{figure}

The field driven polarization switching and the tunneling read current qualify Ferroelectric Tunnel Junctions (FTJs) as high--impedance and low--energy synaptic devices. In this respect, in order to overcome the challenge of ultra--thin ferroelectric layers (1--3 nm) needed in Metal--Ferroelectric--Metal (MFM) stacks, the FTJ design has been recently steered towards Metal--Ferroelectric--Insulator--Metal (MFIM) devices (see Fig.~\ref{Fig:FTJ_sketch}), that are particularly suited for the integration in the BEOL of CMOS circuits \cite{Slesazeck_Nanotechnology2019,Fontanini_ESSDERC2021}.

However, the operation of MFIM FTJs inherently relies on delicate trade--offs. In fact, the depolarization field enables a polarization dependent tunnelling read current, but it also tends to destabilize the polarization in retention and read mode. Moreover, in ferroelectric--dielectric (FE-DE) stacks the charge injection and trapping substantially influence the stability of polarization and its switching dynamics \cite{Park_Nanoscale2021}, as it has also been recognized and debated for ferroelectric--based FETs \cite{Toprasertpong_IEDM2019,Li_VLSI_2020,Deng_IEDM2020}.
Clearly a simulation driven optimization of FTJs is necessary to meet the sometimes conflicting design targets. To this purpose, it is indispensable that a credible modelling framework for the FTJs is developed. This can only be achieved through an appropriate comparison and validation with experiments.

In this work we present an unprecedented joint effort based on an in--house developed numerical model and experiments to understand and to quantitatively investigate the trade--offs implied in the operation of FTJs as synaptic devices. The calibration of the simulator against experiments clarifies some crucial aspects of the device operation related to the charge trapping inside the MFIM stack.
In particular, our results show that either a small or a very large trapping induce, respectively, negligible or complete compensation of the ferroelectric polarization, that hinder the FTJ operation in both cases. In this respect, we report optimal compensation conditions that can be exploited to optimize FTJs as synaptic devices with multiple conductance levels.
 
\section{Device fabrication and experiments}
\label{Sec:Device_Fabrication}

Two types of MFIM FTJ stacks were fabricated with the same nominal thickness for FE and DE films, but through different process sequences. The sketch of the devices is reported in Fig.~\ref{Fig:FTJ_sketch}. The first stack named ‘Stack Anneal--Bottom’ [SAB, Fig.~\ref{Fig:FTJ_sketch}a)] was fabricated as follows: a 30 nm TiN layer was sputtered on $p$$+$$+$ Si substrate as the bottom electrode, then a 3 nm  Al$_2$O$_3$ film followed by a 10 nm Hf$_{0.5}$Zr$_{0.5}$O$_{2}$ (HZO) layer were deposited by atomic layer deposition at 250$^\circ$C. Finally, a 30 nm W layer was deposited as top electrode by sputtering at room temperature. The entire stack underwent a crystallization anneal by RTP at 400$^\circ$C for 120~s in N$_2$ ambient. The top electrodes were patterned by Ti/Pt lift--off followed by W etch in H$_2$O$_2$. A TEM image of a SAB FTJ is shown in Fig.~\ref{Fig:FTJ_sketch}c).

The second FTJ named ‘Post--Anneal Deposition’ [PAD,  Fig.~\ref{Fig:FTJ_sketch}b)] was
fabricated as follows. After the deposition of the 30 nm bottom TiN electrode onto the $p$$+$$+$ Si substrate, a 10 nm HZO layer followed by a 30 nm TiN film were first deposited. The obtained TiN/HZO/TiN stack was RTP annealed at 400$^\circ$C for 120~s in N$_2$ ambient. After the anneal, the top TiN layer was completely etched away and the 3~nm Al$_2$O$_3$ film was deposited at 250$^\circ$C. The deposition and patterning of top W pads was then performed with the process described earlier for the SAB device. Note that, in the PAD FTJ case, the HZO was not in contact with the Al$_2$O$_3$ layer during annealing, while in the SAB stack the HZO and Al$_2$O$_3$ layers were annealed together. This is expected to lead to different HZO/Al$_2$O$_3$ interfaces in the two devices and, thus, to different chemical and electrical boundary conditions for the switching and stabilization of the HZO polarization. In particular, one may expect
a significantly larger density of traps at the annealed HZO/Al$_2$O$_3$ interface of the SAB stack compared to the PAD case. TiN/HZO/TiN capacitors with a 10 nm HZO layer were also fabricated as a reference.

After fabrication, SAB and PAD FTJs are woken--up through 2000 cycles of a triangular waveform at 1~kHz with 4.5~V amplitude. The number of wake--up cycles corresponds to what is needed to wake--up the reference TiN/HZO/TiN capacitors. 
Then, polarization--voltage (P--V) and current--volage (I--V) loops are measured by applying 1~ms long triangular pulses to the top electrode, with the substrate at ground potential. To avoid the breakdown of devices, the maximum voltage amplitudes used for SAB and PAD FTJs are 5 and 4.5~V, respectively.

\section{Modelling framework and calibration}
\label{model}

For the FTJ simulations, we used the in--house numerical model whose details can be found in \cite{Rollo_Nanoscale2020,FontaniniEDTM21,JEDS_21}. In particular, the FE domain dynamics is solved through multi--domain Landau, Ginzburg, Devonshire (LGD) equations, accounting for the 3D electrostatics \cite{Rollo_Nanoscale2020}, whose only free parameters are the LGD constants $\alpha_i$, $\beta_i$, $\gamma_i$, the resistivity $\rho$
of the FE dynamics (setting the time scale), and the domain wall coupling constant $k$. In all simulations, we used $\rho=100\ \Omega\cdot$m, consistently with the values recently reported for HZO based capacitors \cite{Kobayashi_IEDM2016,Kim_IEDM2020,Esseni_Nanoscale2021}, and set $k\simeq0$ in accordance with recent first principle calculations \cite{Ref42_Nanoscale2021} and a detailed comparison between our model and transient operation of ferroelectric based devices \cite{Esseni_Nanoscale2021}.

\begin{figure}[!t]
\centerline{\includegraphics[width=6cm]{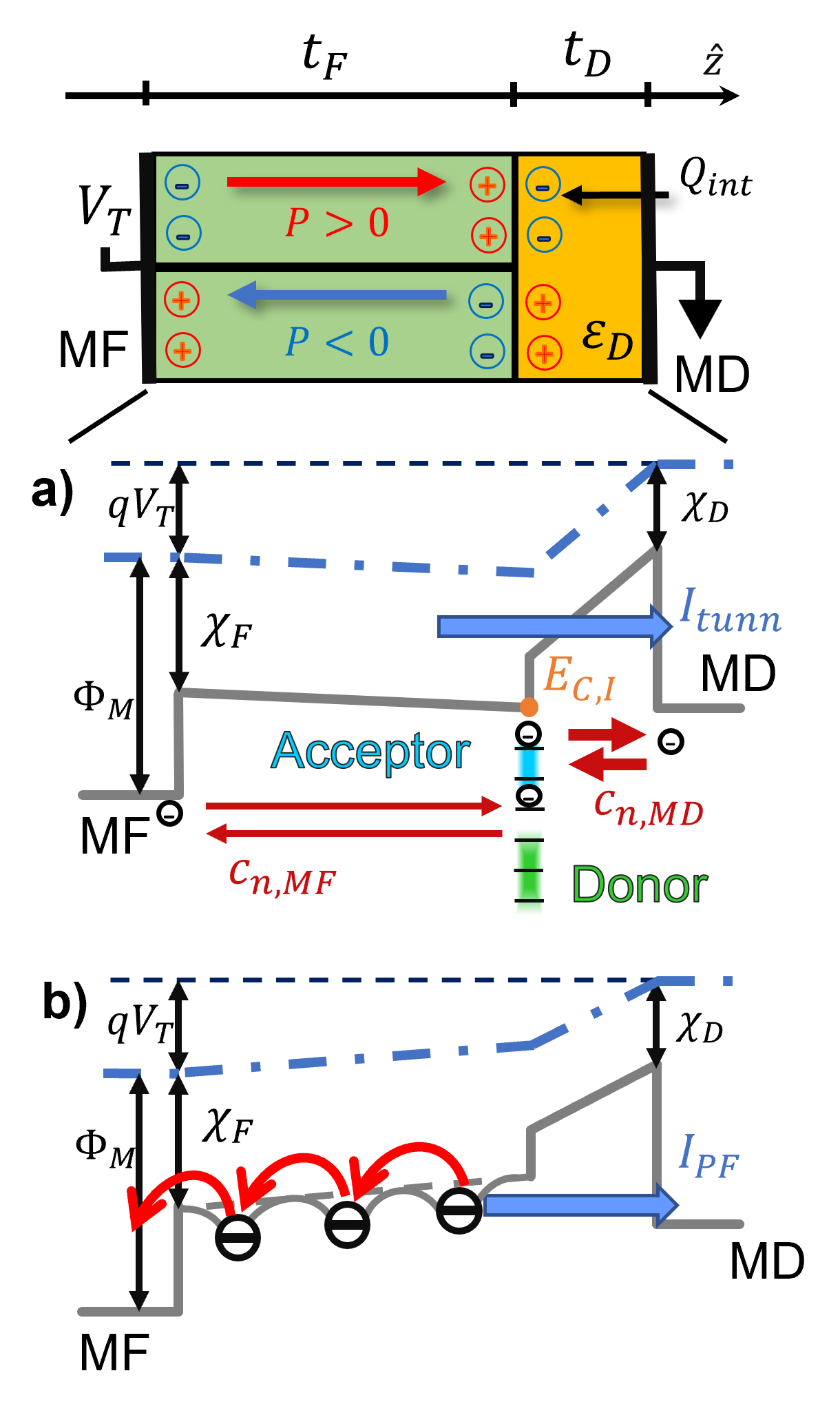}}
\caption{Examples for the band diagram across the FE--DE stack in read mode. Polarization is taken as positive when it points toward the DE--FE interface. $V_T$ is the external applied voltage at the metal contacting the ferroelectric. $\Phi_{M}$ is the electrodes work function. $\chi_{F}=2.4$~eV, $\chi_{D}=1.6$~eV are the electron affinity of the HZO and Al$_{2}$O$_{3}$ layer, $E_{f,MD}$, $E_{f,MF}$ are the Fermi levels of the MD and MF electrodes. (a) The voltage drop $V_D$ in the DE is such that the HZO conduction band minimum, $E_{C,I}$, at the FE--DE interface is smaller than the Fermi level at the MD electrodeI, thus enabling a tunnelling injection, $I_{tunn}$, limited by the dielectric; (b) A smaller $V_D$ compared to results in (a) is expected in case of a Poole--Frenkel current, $I_{PF}$, in shallow HZO traps \cite{Schroeder_JAP2015}.}
\label{Fig:Traps}
\end{figure}

Concerning the read current, \IR, it is assumed to be dominated by the tunnelling across the Al$_2$O$_3$ layer, even if additional mechanisms assisted by defects are also possible \cite{Schroeder_JAP2015}. In this respect, Fig.~\ref{Fig:Traps} qualitatively shows that a similar band bending across the Al$_2$O$_3$ layer is necessary for either a current dominated by tunnelling ($I_{tunn}$, a) or by a Poole--Frenkel mechanism in shallow HZO traps ($I_{PF}$, b).

The model also includes charge injection and trapping, which is described in terms of areal densities at the FE--DE interface \cite{JEDS_21}. However, it is acknowledged that the trap concentrations of this work should be considered as effective areal densities including also a possible charge trapping inside the FE and DE films. The model solves the trap occupation $f_T$ through first order dynamic equations for each trap energy level $E_T$ at the FE--DE surface, assuming a
Fermi occupation functions in the metals MF and MD, respectively at the ferroelectric and dielectric side (see Fig.~\ref{Fig:Traps}) \cite{JEDS_21}.
A detailed balance condition ensures that the $f_T$ is given by the Fermi function at equilibrium (i.e. in steady state at $V_T=0$~V).
As discussed above, the capture rates $c_{MD0}$, $c_{MF0}$ from the MD and the MF electrodes, respectively, are attributed to tunnelling and depend on the area ($\sigma_T$) and energy ($\sigma_E$) cross sections of the traps. 
In each simulation domain the charges $Q_{acc}$, $Q_{don}$ in acceptor and donor traps, respectively, depend on $f_T$ and are proportional to the trap densities $N_{acc}$, $N_{don}$. 
The trap dynamics are solved in all domains self--consistently with the LGD equations for the ferroelectric behaviour \cite{JEDS_21}.

\begin{figure}[!t]
\centerline{\includegraphics[width=7cm]{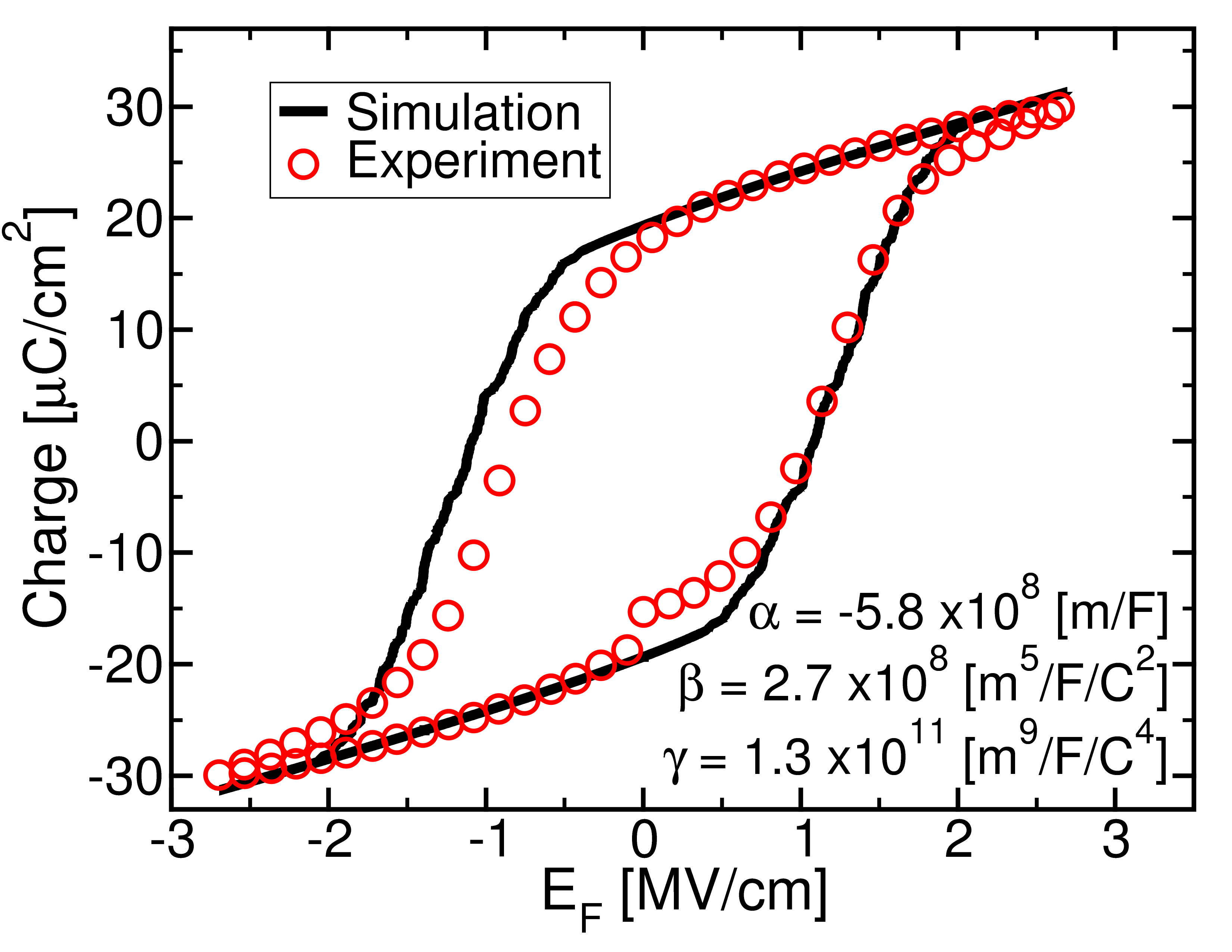}}
\caption{Simulated and experimental charge vs. electric field characteristic for a TiN/HZO/TiN device. Mean values of the gaussian distributed LGD constants are reported. The model reproduces well the experiment, validating also the HZO permittivity used in our calculations.}
\label{Fig:MFM}
\end{figure}

We calibrated the model against experiments. 
In Fig. \ref{Fig:MFM} the simulations based on the LGD equations agree fairly well with the experimental charge versus electric field, $E_F$, curve for a TiN$/$HZO$/$TiN structure, except for the slight asymmetry in the measured P–V. This could be due to differences in top and bottom TiN–HZO interfaces, that are not considered in the simulations.
Such comparison allowed us to extract the LGD constants. In particular, in order to reproduce the experiments well, we considered domain--to--domain variations of the LGD parameters by assuming a normal distribution of $\alpha$, $\beta$ and $\gamma$. The mean values of such distributions are reported in Fig.~\ref{Fig:MFM},
while the standard deviation of each constant are $\sigma_\alpha = 25\% $, $\sigma_{\beta} = 5\% $ and $\sigma_{\gamma} = 8\%$. Throughout this work, we will use these nominal values, if not otherwise stated. 

In all simulations, the work functions of W and TiN electrodes were taken as 4.5 eV, while for the tunneling mass and the permittivity we assumed $m_{FE} = 0.38 \cdot m_{0}$ and $\varepsilon_{FE} = 34 \cdot \varepsilon_{0}$ for the HZO, $m_{DE} = 0.15 \cdot m_{0}$ and $\varepsilon_{DE} = 10 \cdot \varepsilon_{0}$ for the Al$_2$O$_3$, with $m_0$ being the free electron mass and $\varepsilon_0$ the permittivity of vacuum. The values for these material parameters are empirically validated by the fairly good agreement between simulations and experiments.

\section{Evidence of trap contribution in FTJ operation}

\begin{figure}[!t]
\centerline{\includegraphics[width=7cm]{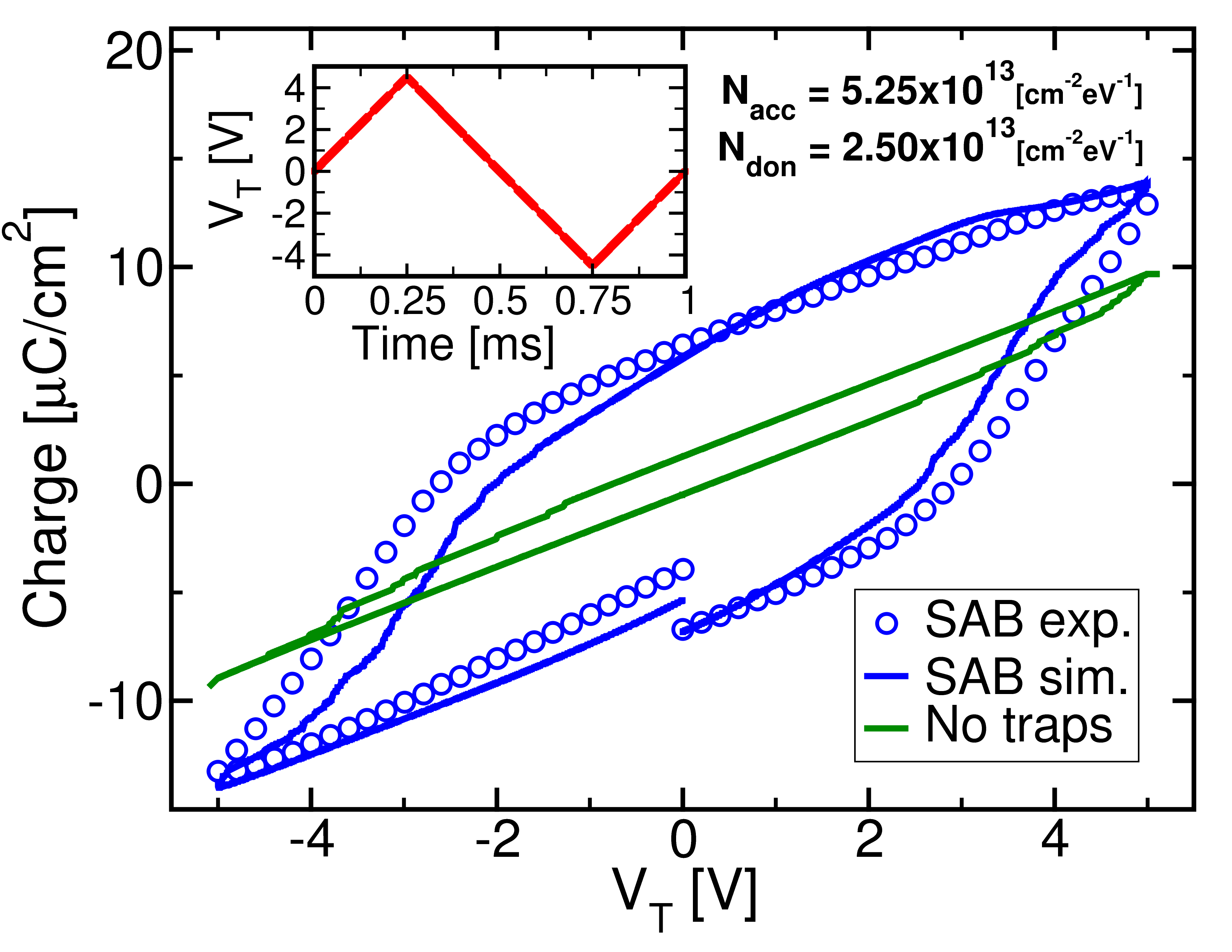}}
\caption{$P$--$V$ characteristics for SAB FTJs (blue circles) measured by using the triangular waveform shown in the inset; Al$_2$O$_3$ thickness is $t_D=3$~nm. Corresponding simulations for no trapped charge (green line) or by accounting for acceptor and donor traps (blue line) are shown.}
\label{Fig:PV_3nm}
\end{figure}

\begin{figure}[!t]
\centerline{\includegraphics[width=7cm]{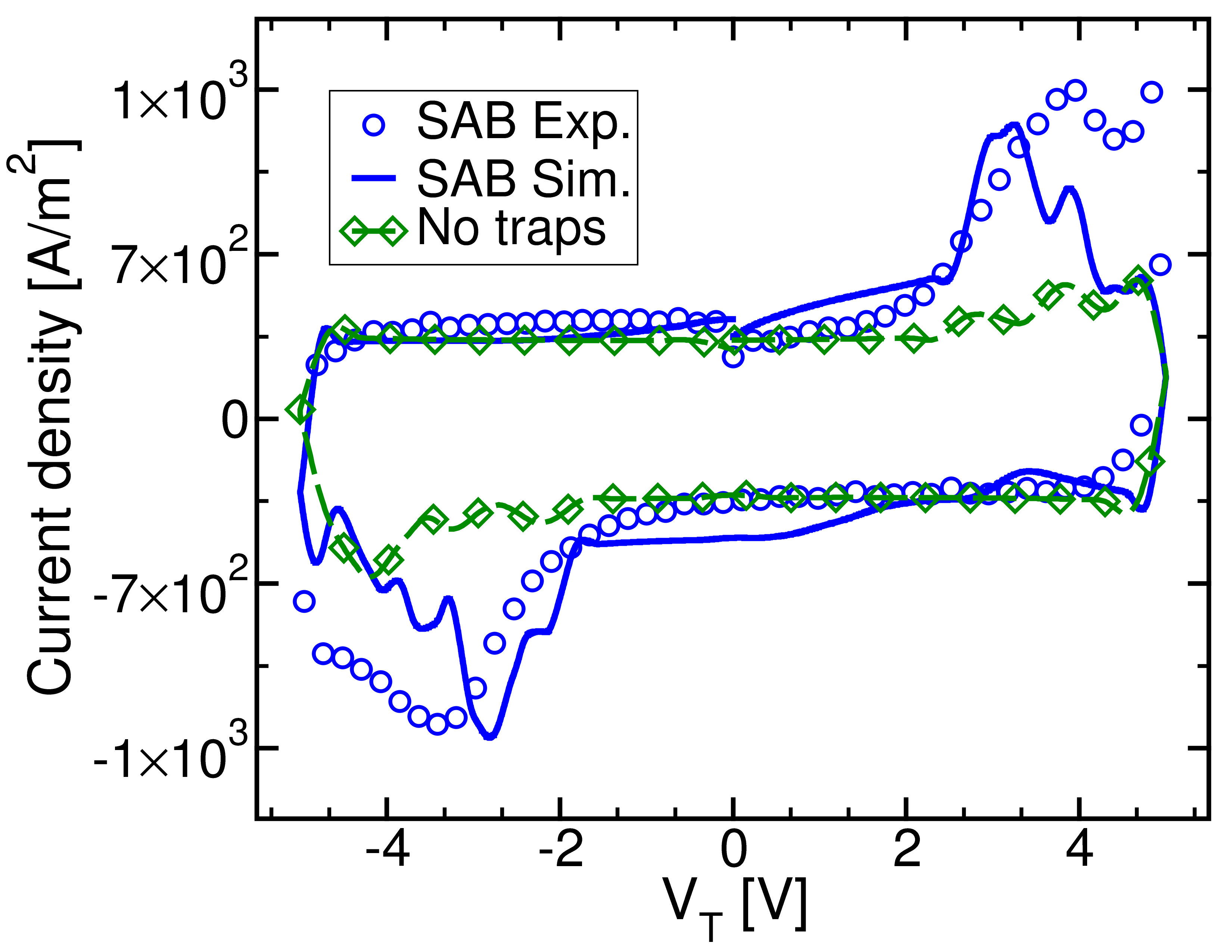}}
\caption{$I$--$V$ curves corresponding
to the $P$--$V$ plots in Fig.~\ref{Fig:PV_3nm}. Trap densities are the same as in Fig.~\ref{Fig:PV_3nm} and are summarized in Tab.~\ref{Tab:Traps_Param}.}
\label{Fig:IV_3nm}
\end{figure}

Figure~\ref{Fig:PV_3nm} compares simulations and experiments concerning the polarization--voltage ($P$--$V$) curves of the SAB devices. As it can be seen, simulations neglecting charge injection and trapping  (green line) exhibit a much more stretched $P$--$V$ curves compared to experiments. These simulation results are
actually qualitatively consistent with experiments reported for a HZO capacitor serially connected to a discrete ceramic capacitor ensuring zero charge injection \cite{Park_Nanoscale2021,Hoffmann_AdvFunctionalMaterials_2021}.
Similarly, in Fig.~\ref{Fig:IV_3nm}, also current--voltage ($I$--$V$) simulations hardly show any switching current when trapping is neglected. 

\begin{figure}[!t]
\centerline{\includegraphics[width=7cm]{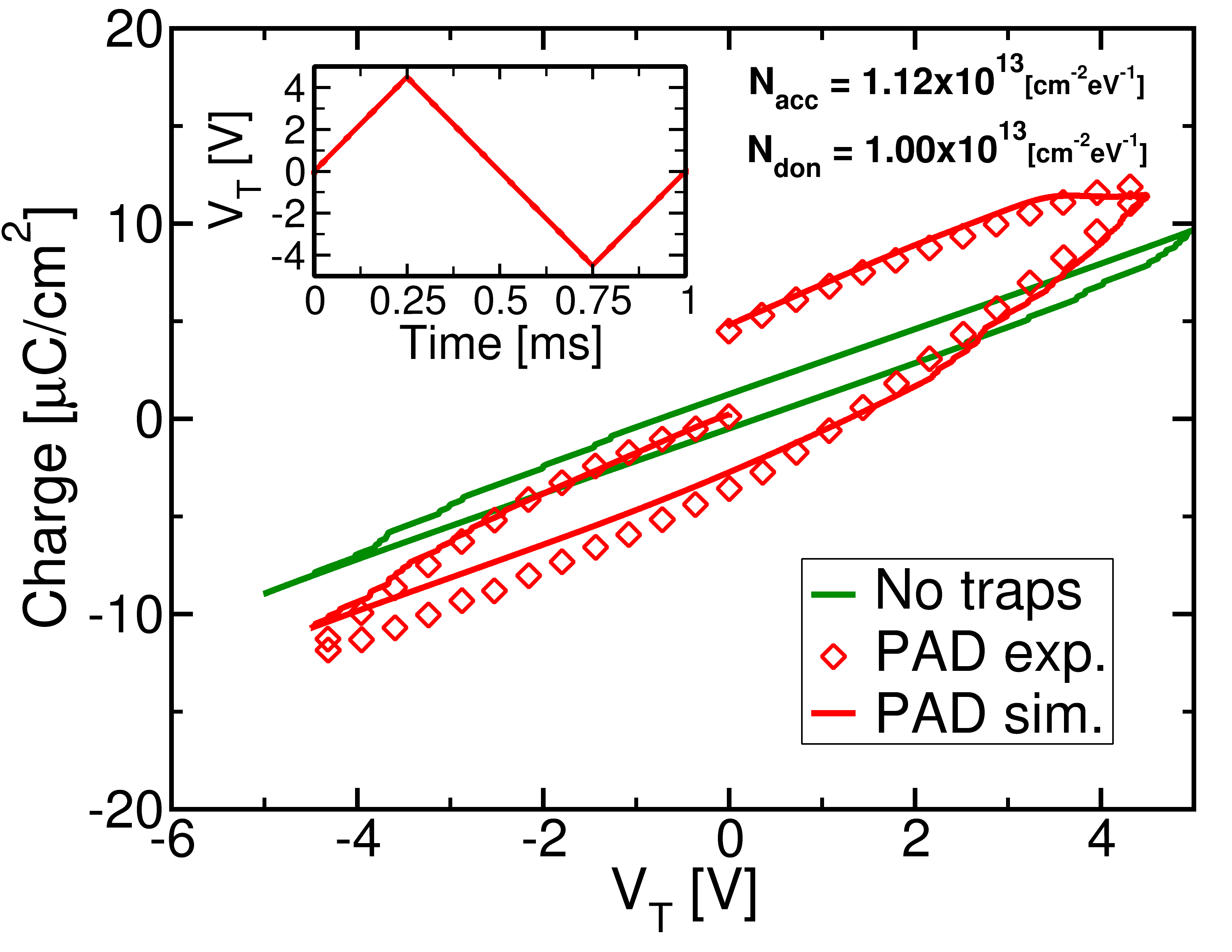}}
\caption{Comparison similar that in Fig.~\ref{Fig:PV_3nm} between simulations and experiments for the $P$--$V$ curves for PAD devices. Corresponding trap densities are reported in Tab.~\ref{Tab:Traps_Param}. Green line shows the no traps case, while red line is the simulation with traps included.}
\label{Fig:PV_PAD_3nm}
\vspace{-5mm}
\end{figure}

A fairly good agreement with $P$--$V$ and $I$--$V$ experiments can instead be achieved by including, in the simulations, an adequate charge density at the FE--DE interface (see blue lines in Figs.~\ref{Fig:PV_3nm} and \ref{Fig:IV_3nm}). This is a clear evidence of the contribution of the charge trapping in the operation of FTJs, that is in line with recent results reported for FeFET devices \cite{Toprasertpong_IEDM2019,Li_VLSI_2020,Deng_IEDM2020}. It is worth mentioning that in Figs.~\ref{Fig:PV_3nm}, \ref{Fig:IV_3nm} and \ref{Fig:PV_PAD_3nm} the LGD constants had to be changed compared to Fig.~\ref{Fig:MFM}, in order to increase the coercive field $E_c$, thus improving the agreement with experiments. This increase of the apparent $E_c$ in MFIM compared to MFM structures has been previously reported and it has been ascribed  to the division of external voltage between DE and FE layer. The capacitance of the ferroelectric increases as the field approaches $E_c$, causing an increased drop across the DE layer and thus higher apparent $E_c$ than in MFM \cite{Jiang_2009}.

\begin{table}[!h]
\caption{Trapping cross sections $\sigma_{T,acc}$, $\sigma_{T,don}$, $\sigma_{E}$ used in simulations throughout the paper and trap densities used in the simulations of Fig.~\ref{Fig:PV_3nm} (SAB) and Fig.~\ref{Fig:PV_PAD_3nm} (PAD).}
\setlength{\tabcolsep}{3pt}
\begin{tabular}{|l|c|c|c|c|c|}
					\hline
					& $N_{acc} $	& $N_{don}$	& $\sigma_{T,acc}$ & $\sigma_{T,don} $	& $\sigma_E $ \\
					& $[cm^{-2}eV^{-1}]$	& $[cm^{-2}eV^{-1}]$	& $[cm^{2}]$ & $[cm^{2}]$	& $[eV]$ \\
					\hline
					\hline
					SAB 	&  	$5.25\cdot10^{13}$		& $2.5\cdot10^{13}$ 		& $3.5\cdot10^{-14}$ 	&	$8\cdot10^{-16}$	& $7\cdot10^{-3}$	\\
					PAD 	&   $1.12\cdot10^{13}$ 	& $1\cdot10^{13}$ 			& $7\cdot10^{-14}$  	&	$7\cdot10^{-14}$ 	& $7\cdot10^{-3}$  \\
					\hline
\end{tabular}
\label{Tab:Traps_Param}
\end{table}

\begin{figure}[!t]
\centerline{\includegraphics[width=7cm]{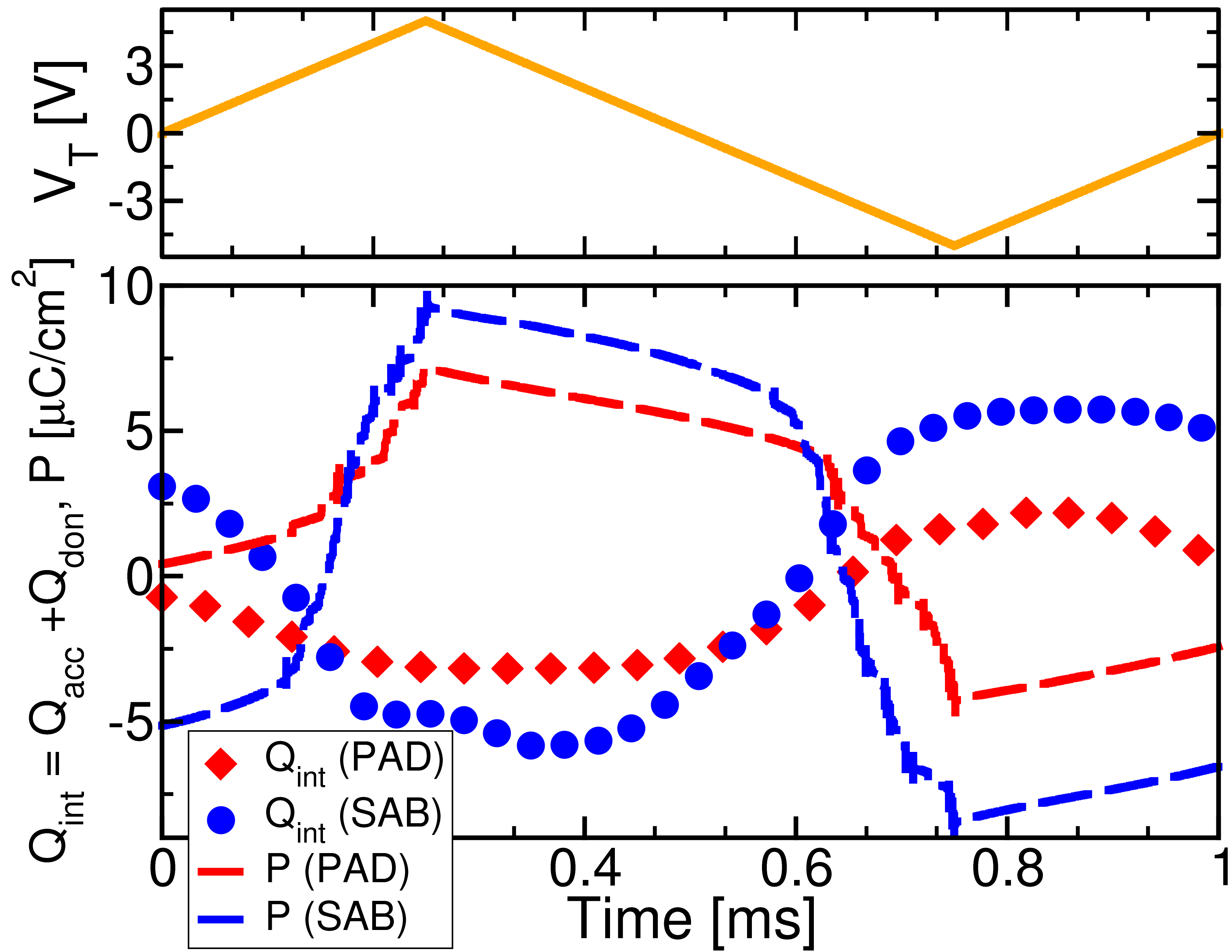}}
\caption{Simulated polarization ($P$, dashed lines) and trapped interface charge ($Q_{int}$, symbols) averaged across the device area for the SAB and PAD simulations in Figs.~\ref{Fig:PV_3nm} and \ref{Fig:PV_PAD_3nm}; the $V_T$ waveform is shown as a yellow solid line. The used interfacial trap densities are reported in Tab.~\ref{Tab:Traps_Param}.}
\label{Fig:Qint}
\end{figure}

Figure~\ref{Fig:PV_PAD_3nm} reports an analysis similar to that in Fig.~\ref{Fig:PV_3nm} for the PAD devices. Even for the PAD FTJs, the simulations neglecting trapping (green line) exhibit large discrepancies with experiments (symbols). However, the trap density matching simulations with experiments is smaller than the concentration used for the SAB devices (see Tab.~\ref{Tab:Traps_Param}). 
It should be mentioned that not only the processing conditions but also the FTJ wake--up may influence the density of the electrically active defects. However, since both FTJs underwent the same wake--up sequence,
we expect that this difference is due to the different annealing conditions for the two devices. The difference in the simulated interfacial charge in SAB and PAD samples is best illustrated by Fig.~\ref{Fig:Qint}, reporting the average polarization and trapped charge $Q_{int}=(Q_{acc}+Q_{don})$ along a triangular $V_T$ waveform. 
Figure~\ref{Fig:Qint} shows that the polarization is partly compensated by $Q_{int}$ in both devices, but the effect is more prominent in SAB FTJs. Again, this is ascribed to the different annealing sequence between the two devices and Fig.~\ref{Fig:Qint} provides an insight about the role of the trap density at the HZO/Al$_2$O$_3$ interface in the FTJ operation.
At the present time we are not able to explore additional annealing conditions besides those employed for SAB and PAD devices. However, the conditions explored here are in the suitable range for the FTJ integration in a CMOS BEOL. This is an important aspect for the neuromorphic applications of FTJs, so that the paper is actually focused on BEOL compatible process conditions.

\begin{figure}[!t]
\centerline{\includegraphics[width=7cm]{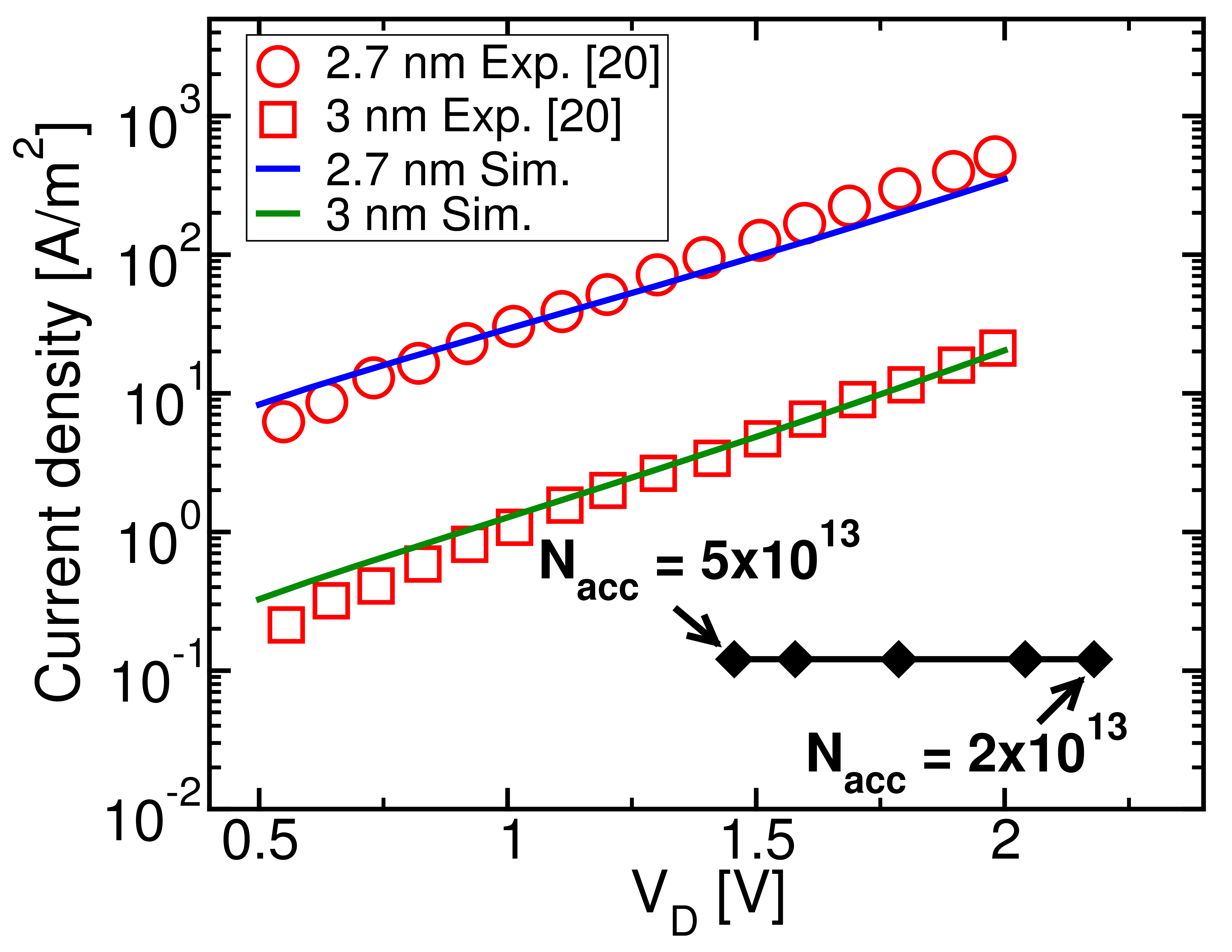}}
\caption{Measured current density for a SAB device at a read voltage $V_R=2$~V (black diamonds) versus the corresponding voltage drop, $V_D$, across the 3~nm Al$_2$O$_3$ estimated by using simulations with different density, $N_{acc}$ [cm$^{-2}$eV$^{-1}$], of acceptor traps.
In these simulations the density of donors does not affect the final result.
Measured tunnelling current in thin SiO$_2$ layers (red symbols) and corresponding simulations with the tunnelling model of this work are reported as a reference.}
\label{Fig:Jtun}
\end{figure}

In read mode, the compensation of the positive polarization is expected to reduce the band bending and thus the read current $I_R$ (see Fig.~\ref{Fig:Traps}). In this regard, Fig.~\ref{Fig:Jtun} shows the measured ON current density of about 0.12~A/m$^2$ (black diamonds) for SAB FTJs, together with the tunnelling current density in metal--SiO$_2$--metal (MIM) systems \cite{Matsuo_JAP1999}; incidentally, our tunnelling model is in good agreement with experiments in the MIM systems. For the FTJ device, the voltage drop, $V_D$, across the Al$_2$O$_3$ layer is estimated by simulations and for different $N_{acc}$.
Figure~\ref{Fig:Jtun} shows that the estimated $V_D$ for $N_{acc}\simeq 5 \cdot 10^{13}$~cm$^{-2}$eV$^{-1}$ (see SAB parameters in Tab.~\ref{Tab:Traps_Param}) is clearly smaller than the value $V_D=(\Phi_{M}-\chi_{F})/q=2.1$~V required to have a read current $I_R$ limited by the $I_{tunn}$ through the Al$_2$O$_3$ film alone [see Fig.~\ref{Fig:Traps}a)]. Consistently with this picture, the measured $I_R$ is much smaller than the $I_{tunn}$ in MIM systems, and it may be limited by Poole--Frenkel conduction [see $I_{PF}$ in Fig. \ref{Fig:Traps}b)].

\section{Polarization--compensation--aware design of the FTJ}
\label{sec:Design_FTJs}

\begin{figure}[!t]
\centerline{\includegraphics[width=8cm]{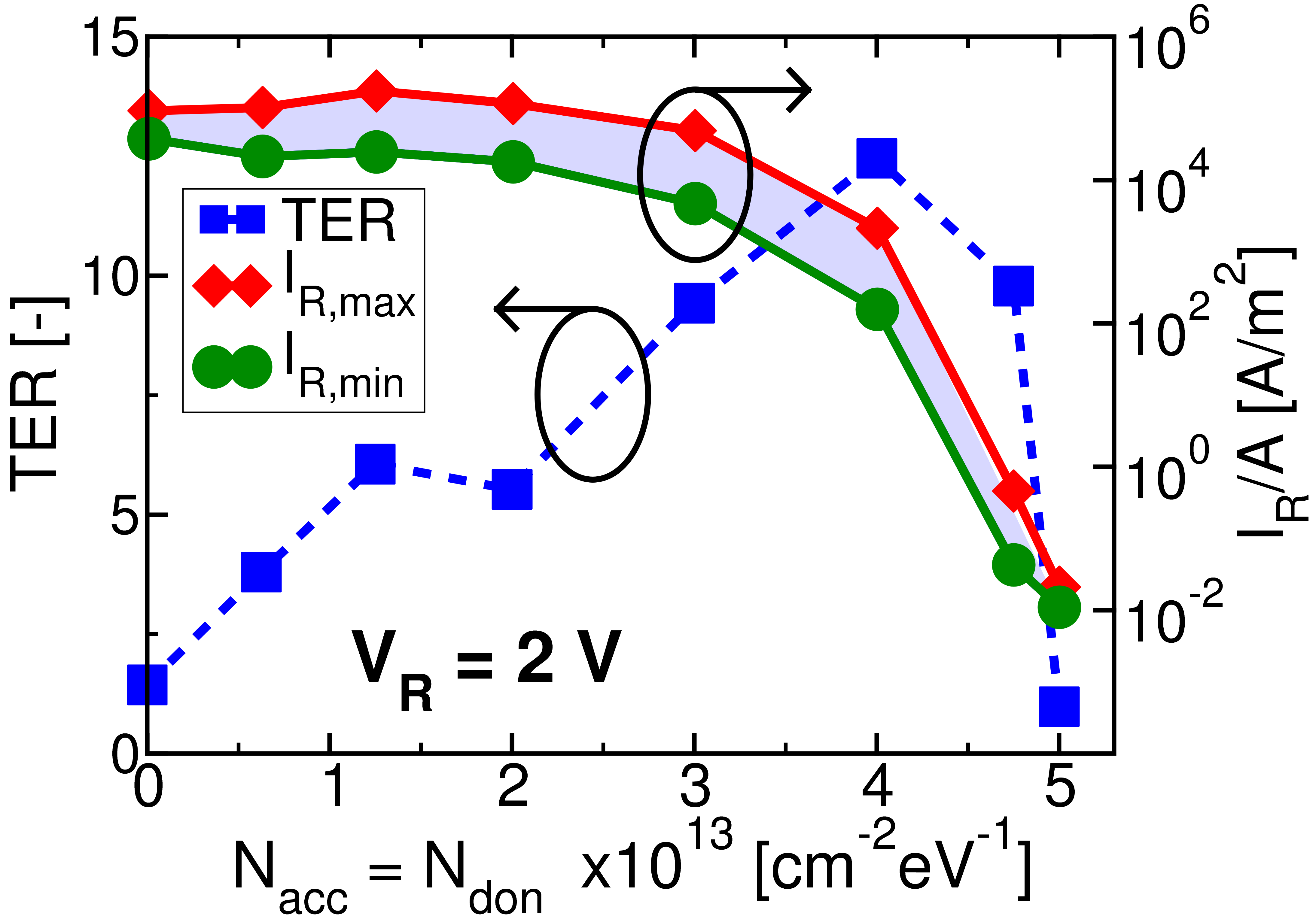}}
\caption{Minimum, $I_{R,min}$, and maximum, $I_{R,max}$, simulated read current (right $y$--axis) versus the trap density at read voltage $V_R=2$~V. $I_{R,min}$ and $I_{R,max}$ correspond to a SET pulse voltage of $V_{SET, min}=2.5$~V and $V_{SET, max}=4.5$~V, respectively. The corresponding tunnel electroresistance, TER~$=(I_{R, max}-I_{R, min})/I_{R, min}$ is reported (left $y$--axis).}
\label{Fig:jter}
\end{figure}

In recent designs of neuromorphic processors, a reasonable target for the minimum read current is set to about 100~pA \cite{Qiao_BioCAS_2016,Sharifshazileh_ICECS2018}, which requires a read current density of about $I_R/A\simeq100$~A/m$^2$ for an FTJ area of $A \approx 1$~$\mu$m$^2$. Hereafter, we use our calibrated simulations to examine the optimal design of FTJs in terms of $I_R$ and tunneling electroresistance TER~$=(I_{R,max}-I_{R,min})/I_{R,min}$. For the simulations in this section, the LGD constants are those in Fig.~\ref{Fig:MFM}, calibrated on the MFM device, in order to precisely account for the characteristics of the integrated HZO layer. 

\begin{figure}[!t]
\centerline{\includegraphics[width=7cm]{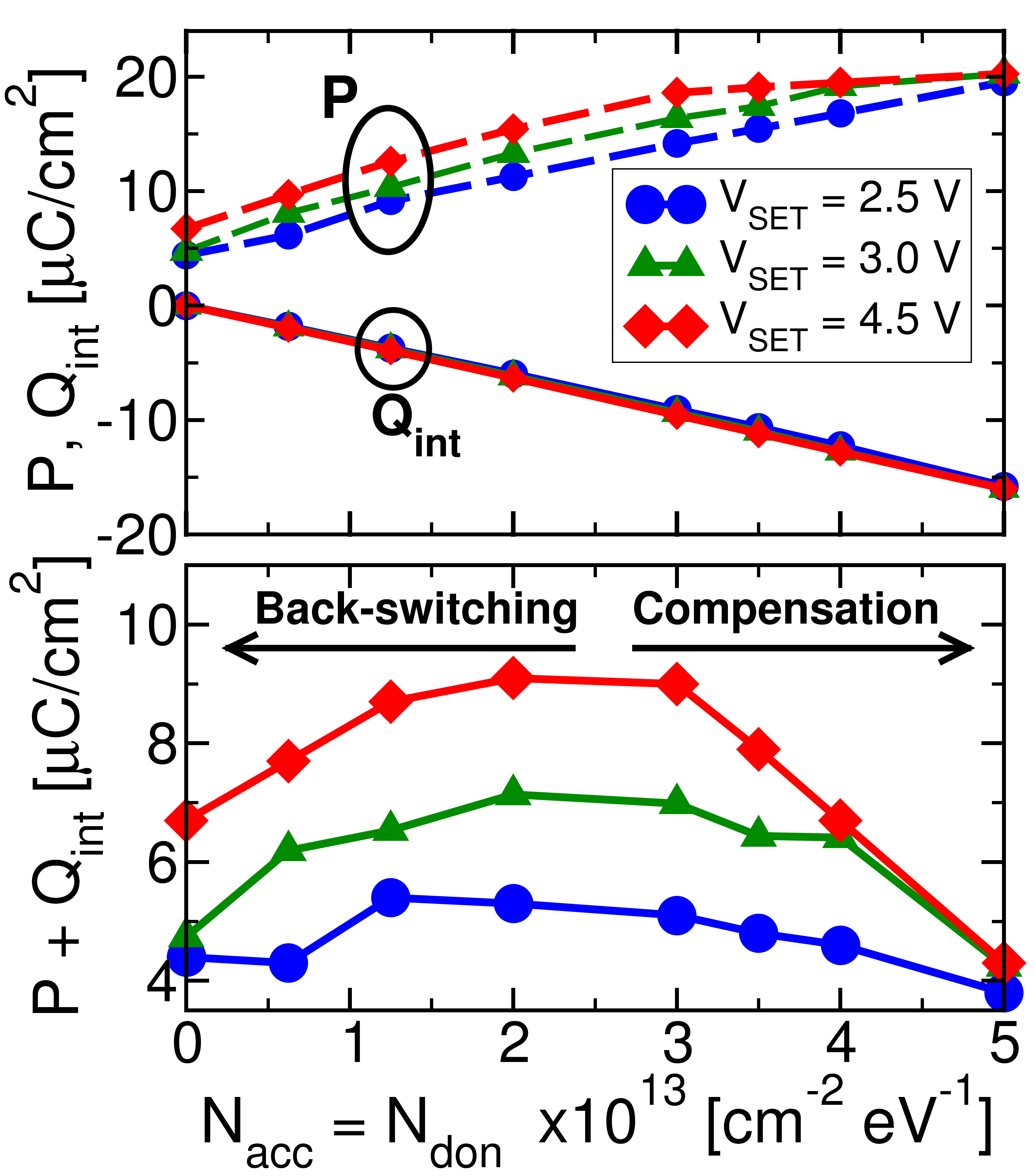}}
\caption{Average polarization, $P$, interface charge, $Q_{int}$, and net or compensated polarization $(P+Q_{int})$ vs. the trap density in read condition ($V_R=2$~V) and for different SET voltages.
At low $N_{acc}$,  $(P+Q_{int})$ degrades due to depolarization field and back--switching, while at large $N_{acc}$,  $(P+Q_{int})$ decreases due to a large $P$ compensation.}
\label{Fig:avgcomp}
\end{figure}

\begin{figure}[!t]
\centerline{\includegraphics[width=7cm]{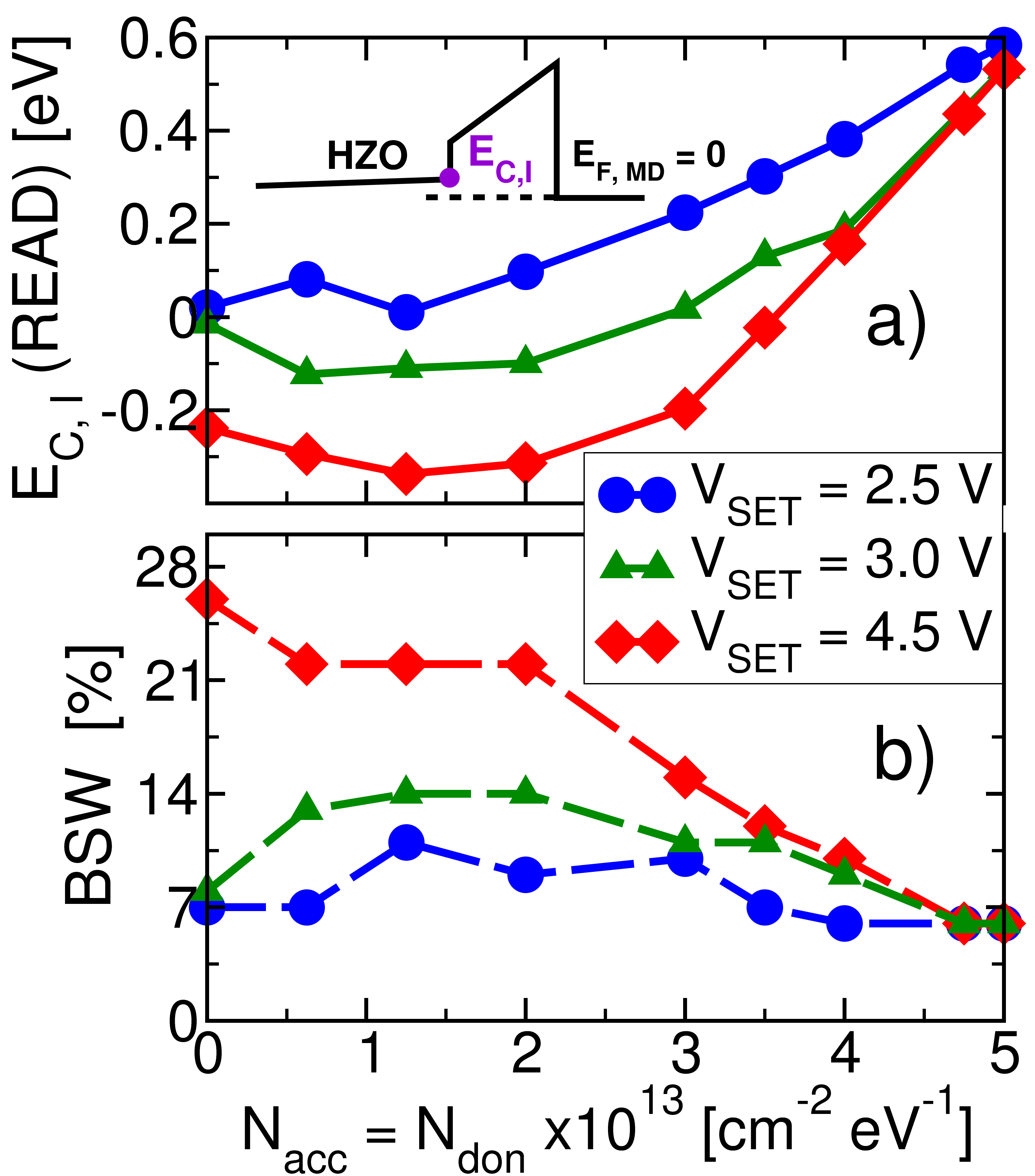}}
\caption{a) HZO conduction band minimum, $E_{C,I}$, in read mode  versus the trap density, where the Fermi level at the MD electrode is $E_{F,MD}=0$~eV (see inset).
b) Ferroelectric domain back--switching, BSW, defined as the percentage of domains having a negative $P$ in retention after being set to a positive $P$ during the SET pulse.}
\label{Fig:ecinacc}
\end{figure}

Figure~\ref{Fig:jter} reports the simulated minimum and maximum read current densities and TER for a FTJ having an Al$_2$O$_3$ thickness scaled down to $t_D=2.5$~nm and for different trap densities $N_{acc}=N_{don}$\footnote{This simplifying assumption of an equal density of acceptor and donor traps has been introduced in order to limit the number of free parameters in the analysis carried out in this section. We expect that the main outcome of this discussion does not change if $N_{acc}$ and $N_{don}$ are allowed to be different.}. Despite the uncertainties that admittedly affect our calculations of the read current and have been also discussed in Fig.~\ref{Fig:Jtun}, the $I_R$ and the TER values calculated at large trap densities are in--line with recent literature on SoA FTJ devices, showing read currents in the 0.05--0.6~A/m$^2$ range and TER values between 5 and 12.5 \cite{Max_JEDS2019,Ryu_Nature2019,Hur20}.

Quite interestingly, Figure~\ref{Fig:jter} shows that the TER exhibits an optimal value for $N_{acc}$ around $4\cdot10^{13}$ cm$^{-2}$eV$^{-1}$, which stems from the behavior of the net or compensated polarization $(P+Q_{int})$ illustrated in Fig.~\ref{Fig:avgcomp}.
In fact, at low $N_{acc}$ values, the compensation of the positive $P$ is too weak, so that the depolarization field $E_F$ increases and destabilizes the polarization. 
In this respect, we calculated the difference between the number of domains with positive spontaneous polarization $P$ at the maximum value of the SET voltage and at zero external bias (i.e. during retention) and then we divided this value by the overall number of domains. This is indeed the fraction of the domains set to positive $P$ during the SET pulse that back--switch to negative $P$ during retention. We defined such value as back–-switching in Fig.~\ref{Fig:ecinacc}b), which shows that at low $N_{acc}$ a significant back--switching of the ferroelectric domains occurs. 
The large back–-switching at low $N_{acc}$ deteriorates the TER in Fig.~\ref{Fig:jter}.

At large $N_{acc}$, instead, there is an excessive compensation that reduces $(P+Q_{int})$. This implies an increase of the HZO conduction band minimum, $E_{C,I}$, at the FE--DE interface during the read mode, which is reported in Fig.~\ref{Fig:ecinacc}a). Such an $E_{C,I}$ raise leads to the large $I_R$ drop at high $N_{acc}$ shown in Fig.~\ref{Fig:jter}. Moreover, a large compensation tends to merge the states obtained by different $V_{SET}$ values, which seriously hinders a multi--level read current and thus the operation as a synaptic device.

\begin{figure}[!t]
\centerline{\includegraphics[width=7cm]{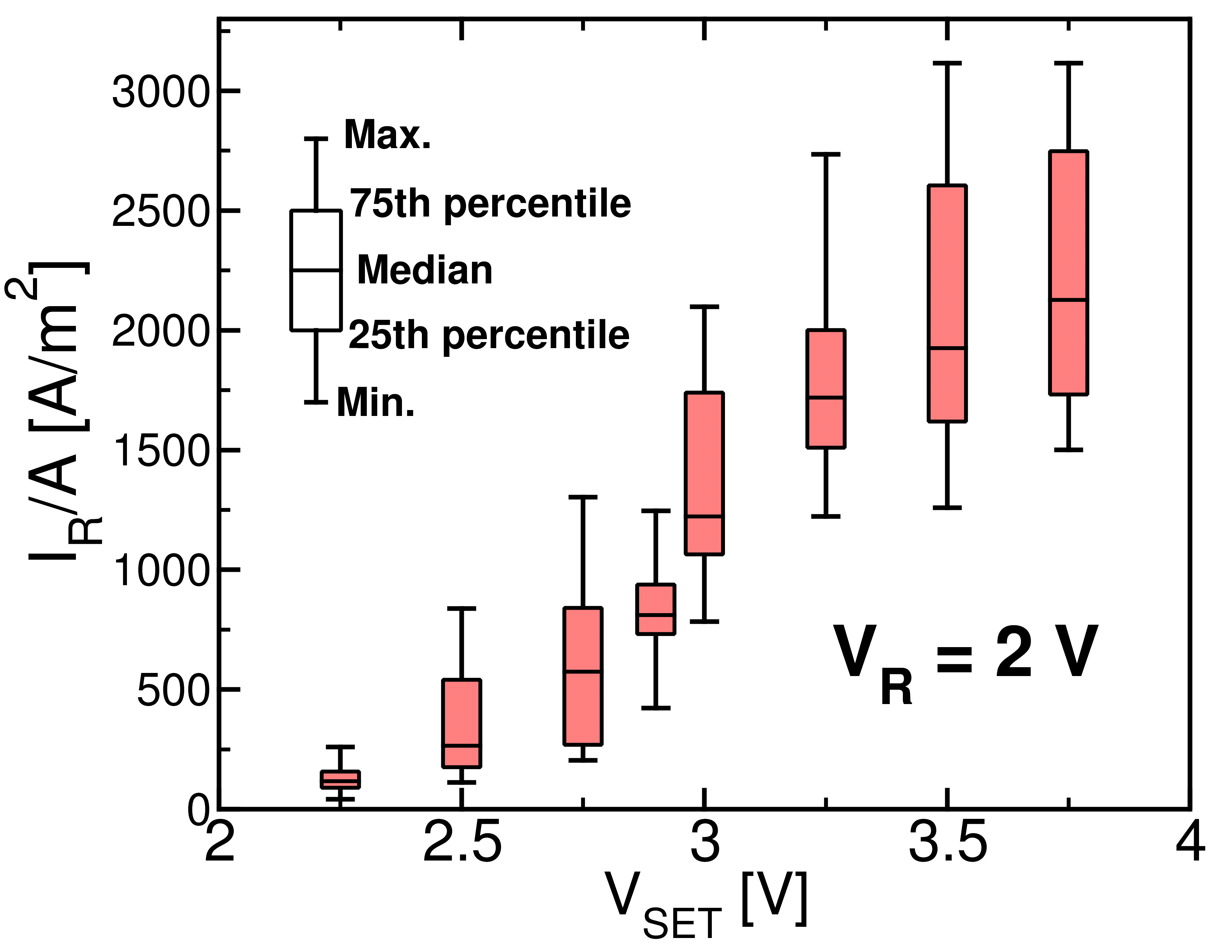}}
\caption{Simulation results for the FTJ of Fig.~\ref{Fig:jter} under optimal compensation conditions corresponding to $N_{acc}=4\cdot10^{13}$~cm$^{-2}$eV$^{-1}$. The box plots reveal overlapping distributions for the current levels for small area FTJs with $n_D=100$ ferroelectric domains. Simulations have been performed for a small device area of 2500~nm$^2$ (corresponding to a domain size $d=5$~nm).}
\label{Fig:levels}
\end{figure}

The results of Fig.~\ref{Fig:jter} and the inspection of the internal quantities performed in Figs.~\ref{Fig:avgcomp} and \ref{Fig:ecinacc}, clearly show that an optimal compensation condition exists for the operation of FTJs.
In this respect, as discussed above, both negligible and excessive charge--trapping--induced compensation of the ferroelectric polarization are detrimental for the operation of the FTJs under study. Indeed, there exists a quite delicate balance between the suppression of back--switching and TER values enabling a multi--level operation in read mode.
In Fig.~\ref{Fig:levels}, for a given FTJ structure, we leverage the optimal design condition in Fig.~\ref{Fig:jter} and show that eight current levels can be placed in the memory window of the FTJ for the optimal compensation condition corresponding to $N_{acc}=4\cdot10^{13}$~cm$^{-2}$eV$^{-1}$, thus enabling a 3--bit synaptic weight resolution. 
It is worth mentioning that such an optimal condition corresponds to a $V_D$ across the Al$_2$O$_3$ layer that is very close to or larger than the value $(\Phi_{M}-\chi_{F})/q=2.1$~V required to have 
tunnelling through the thin dielectric film alone [see Figs.~\ref{Fig:Traps}a) and \ref{Fig:ecinacc}].
A comparison of the calculated $I_R$ with experiments suggests that it is probably difficult in actual FTJs to reach this favourable condition where $I_R$ is only limited by the tunnelling through the Al$_2$O$_3$ layer, as also discussed in Fig.~\ref{Fig:Jtun}.


\section{Conclusions}
We presented a joint effort between numerical modelling and experiments to demonstrate the crucial role of polarization compensation due to trapped charges in HZO based FTJs. We identified and physically explained the optimal compensation condition for a robust operation of FTJs.

Even if a tailoring of trap densities may be challenging from a technological standpoint, we experimentally demonstrated that appropriate processing steps and annealing conditions can lead to different trap densities at the FE--DE interface of fabricated FTJs. 
Moreover, the trap density can be partly controlled in ALD deposited Al$_2$O$_3$ films by tuning the deposition parameters such as the precursor pulse duration, the oxidant precursor type (water, O$_2$ plasma, ozone), the deposition temperature, and the dosing time. These processing parameters have been shown to change the Al$_2$O$_3$ defect density at least by an order of magnitude \cite{Rahman_2020, Kim_JAP2002, Henkel_2017}. Another pathway is to include an ultrathin charge trap layer between the tunneling dielectric and HZO \cite{Wu_EDL2010}.
As a final remark, 
the insertion of an ultra--thin metal layer at the FE--DE interface may also be an interesting design option to control the compensation condition in FTJs, as it has been suggested in \cite{Frank_TED2014}.

\appendices


\section*{Acknowledgment}

The authors thank Dr. A. Hammud, FHI--Berlin, for support.

\bibliographystyle{bibtex/MyIEEEtran.bst}
\bibliography{bibtex/biblio.bib}

\end{document}

%% file: macros.tex
\newcommand{\IR}{\ensuremath{\textrm{I}_{R}}}

